\documentclass[twocolumn,prl]{revtex4}

\usepackage{dcolumn}
\usepackage{amsmath}

\usepackage{graphicx}
\usepackage{rotating}

\begin{document}

\newcommand{\defn}{\textit}
\newcommand{\half}{\mbox{$\frac12$}}
\newcommand{\dd}{{\rm d}}
\newcommand{\ii}{{\rm i}}
\renewcommand{\O}{{\rm O}}
\newcommand{\e}{{\rm e}}
\newcommand{\set}[1]{\lbrace#1\rbrace}
\newcommand{\av}[1]{\langle#1\rangle}
\newcommand{\eref}[1]{(\ref{#1})}
\newcommand{\etal}{{\it{}et~al.}}

\newlength{\figurewidth}
\setlength{\figurewidth}{0.95\columnwidth}
\setlength{\parskip}{0pt}
\setlength{\tabcolsep}{6pt}

\title{The spatial structure of networks}
\author{Michael T. Gastner}
\affiliation{Department of Physics, University of Michigan, Ann Arbor,
MI 48109--1120}
\author{M. E. J. Newman}
\affiliation{Department of Physics, University of Michigan, Ann Arbor,
MI 48109--1120}

\begin{abstract}
We study networks that connect points in geographic space, such as
transportation networks and the Internet.  We find that there are strong
signatures in these networks of topography and use patterns, giving the
networks shapes that are quite distinct from one another and from
non-geographic networks.  We offer an explanation of these differences in
terms of the costs and benefits of transportation and communication, and
give a simple model based on the Monte Carlo optimization of these costs
and benefits that reproduces well the qualitative features of the networks
studied.
\end{abstract}
\maketitle

There has in the last few years been considerable interest within the
physics community in the analysis and modeling of networked systems
including the world wide web, the Internet, and biological, social, and
infrastructure networks~\cite{AB02,DM02,N03}.  Some of these networks, such
as biochemical networks and citation networks, exist only in an abstract
``network space'' where the precise positions of the network nodes have no
particular meaning.  But many others, such as the Internet, live in the
real space of everyday experience, with nodes (e.g.,~computers in the case
of the Internet) having well-defined positions.  Most previous studies have
ignored the geography of networks, concentrating instead on other issues.
Here we argue that geography matters greatly, and to ignore it is to miss
some of these systems' most interesting features.

A network in its simplest form is a set of nodes or \defn{vertices} joined
together in pairs by lines or \defn{edges}.  We consider networks in which
the vertices occupy particular positions in space.  The edges in these
networks are often real physical constructs, such as roads or railway lines
in transportation networks~\cite{SDCSMM03}, optical fiber or other
connections in the Internet~\cite{W88,YJB02}, cables in a power
grid~\cite{WS98}, or oil pipelines~\cite{BHLMB03}.  In other cases the
edges may be more ephemeral, such as flights between
airports~\cite{GMTA03}, business relationships between
companies~\cite{M82}, or wireless communications~\cite{M01}.

Interest in the spatial structure of networks dates back to the economic
geography movement of the 1960s~\cite{G60,HC69} and particularly the work
of Kansky~\cite{K63}.  Early work was hampered however by limited data and
computing resources, and geographers' attention moved on after a while to
other topics.  Networks have come back into the limelight in recent years,
particularly as a result of interest among physicists, but spatial aspects
have not received much attention.  The best known theoretical models of
networks either make no reference to space at all~\cite{MR95,BA99}, or they
place vertices on simple regular lattices whose structure is quite
different from that of real systems~\cite{WS98,Kleinberg00}.  The successes
of these models---which are considerable---have been in their ability to
predict topological measures such as graph diameters, degree distributions,
and clustering coefficients.  Empirical studies of networks, even networks
in which geography plays a pivotal role, have, with some
exceptions~\cite{YJB02,GK04,GMTA04,CS04}, similarly focused almost
exclusively on topology~\cite{FFF99,ASBS00,SDCSMM03}.

In this paper we look at three specific networks, particularly emphasizing
their spatial form.  The three networks are the Internet, a road network,
and a network of passenger flights operated by a major airline.  To make
comparison between the networks easier we limit our studies to the United
States, and we exclude Alaska and Hawaii to avoid problems of disjoint
maps.

The first of our three networks is the Internet.  We examine the network in
which the vertices are autonomous systems (ASes) and the edges are data
connections between them (technically, direct-peering relationships).
The topology of the connections between ASes can be inferred from routing
tables.  In our studies we have made use of the collection of routing
tables compiled by the University of Oregon's Route Views
project~\cite{RV03}.  To determine the geographical parameters of the
network we use NetGeo~\cite{NG03}, a software tool that can return
approximate latitude and longitude for a specified AS.  Combining these two
resources a geographic map of the Internet was created, from which were
then deleted all nodes falling outside the lower 48 states.  This leaves a
network of $7049$ nodes and $13\,831$ edges for data from March 2003.

Our second network is the US interstate highway network in which the
vertices represent intersections, termination points of highways, and
country borders, and the edges represent highways.  Vertex positions and
edges were extracted from GIS databases.  For data from the year 2000 the
network has 935 vertices and 1337 edges.  Our third network, the airline
network, is similarly straightforward.  In this network the vertices
represent airports and there is an edge between every pair of airports
connected by a scheduled flight.  The particular case we study is the
published schedule of flights for Delta Airlines for February 2003, for
which there are 187 vertices and 825 edges.  Geographic locations of
airports were found from standard directories.

\begin{figure}[t]
\centering
\resizebox{8cm}{!}{\includegraphics{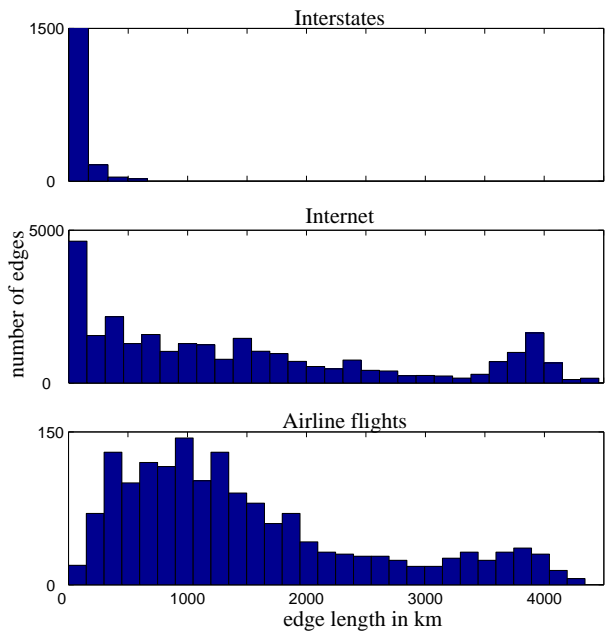}}
\caption{Histograms of the lengths of edges in the three networks studied
here.}
\label{lngthdstr}
\end{figure}

We focus initially in our analysis of these networks on three fundamental
properties: edge lengths, network diameter, and vertex degrees.  In
Fig.~\ref{lngthdstr} we show the distribution of the lengths in kilometers
of edges in each of our networks.  Common to all three networks is a clear
bias towards shorter edges, which is unsurprising since long edges are
presumably more expensive to create and maintain than short ones.  When we
look more closely, however, the networks show some striking differences.
The road network has only very short edges, on the order of 10km to 100km,
while the Internet and airline network have much longer ones.  The latter
two networks also both have bimodal distributions, with a large fraction of
edges of length 2000km or less, and then a smaller but distinct peak of
longer edges around 4000km~\footnote{Although we do not dwell in it in this
paper, the Internet and the airline network do differ at the shortest edge
lengths, the Internet having a strong peak for edges of 100km or less,
while the airline network seems deliberately to avoid such short edges,
having a dip in the distribution at the shortest length scales.  This is
presumably an effect of economic pressures: very short airline flights are
uneconomical because passengers can conveniently drive the same distance
for less money.}.  (These are continent-spanning edges, like coast-to-coast
flights in the airline network.)

Simple Euclidean distance between vertices is not the only measure of
distance in a network however.  Another commonly used measure is the
so-called graph distance, which measures the number of edges traversed
along the shortest path from one vertex to another---the number of ``legs''
of air travel, for instance, or the number of ``hops'' an Internet data
packet would make.  The largest graph distance between any two points in a
network is called the graph diameter, and it varies widely between our
networks.  For the highway network for example the diameter is 61, but it
is just 8 for the Internet, even though the latter network has far more
vertices.  And for the airline network the diameter is only~3.  In the
jargon of the networks literature, the Internet and the airline network
form ``small worlds,'' while the interstate network does not.

Euclidean edge lengths and graph distances are not unrelated: in a graph
like the road network, which is composed mainly of short edges, one will
need to traverse a lot of such edges to make a long journey, so we would
expect the diameter to be large.  Conversely, the presence of even just a
few long edges makes for much smaller diameters, as demonstrated recently
by Watts and Strogatz~\cite{WS98}.  Thus there seems to be a pay-off
between Euclidean distance and number of legs in a journey, an idea that we
exploit below to help explain the observed structure of our networks.

Another way in which our networks differ is in the degrees of their
vertices.  (The degree of a vertex is the number of edges connected to it.)
The highest degree of any vertex in the highway network is~$4$, which means
that the best connected vertex links directly to only $0.4\%$ of other
vertices.  In the airline network by contrast, the maximum degree is 141 or
$76\%$ of the network, while for the Internet it is 2139 or $30\%$.
High-degree vertices that connect to a significant fraction of the rest of
the network are commonly called ``hubs''; the airline network and Internet
thus both contain at least one hub (in fact each contains several), whereas
the road network contains none~\footnote{The existence of hubs in the
airline network is of course well known to travelers, and their existence
in the Internet has also been known for some years~\cite{FFF99}.}.

We would like to understand how the observed structure of our networks is
related to their geographical nature, and the origin of the marked
differences between the networks.  We present two approaches that shed
light on these questions.  The first is empirical in nature, the second
theoretical.

At the empirical level, many of the features we observe in these networks
can be explained in terms of spatial dimension.  Each of our networks is of
course two-dimensional in a geographic sense, since it lives on the
two-dimensional surface of the Earth.  However, one can also ask about the
effective dimension of the network itself~\cite{NW99b}.  We find that, in a
sense we will shortly define, the Internet and airline networks are not
really two-dimensional at all, but the road network is.

The road network is, in fact, almost planar.  That is, it can be drawn on a
map without any edges crossing.  This automatically gives it a
two-dimensional form and helps us to understand why its edges are so short:
if edges are not allowed to cross then they cannot travel far before they
run into one another.  It also goes some way towards explaining the
network's low vertex degrees: it can be proved that the mean
degree~$\bar{k}$ of a planar graph is strictly less than~6~\cite{West96}
and indeed we find that the mean degree of the road network is
$\bar{k}=2.86$.  For the airline network on the other hand $\bar{k}=8.82$,
so this network cannot be planar.  This is not an entirely persuasive
argument however.  The Internet has mean degree $\bar{k}=3.93$, which is
not large enough to rule out planarity, and the highway network is actually
not perfectly planar, having a small number of road crossings so that
rigorous demonstrations of planarity such as Kuratowski's
theorem~\cite{West96} or the Hopcroft--Tarjan planarity
algorithm~\cite{HT74} fail.  We would like, therefore, some other more
flexible way of probing the dimension of our networks.  We propose the
following.

\begin{figure}
\begin{center}
\resizebox{\columnwidth}{!}{\includegraphics{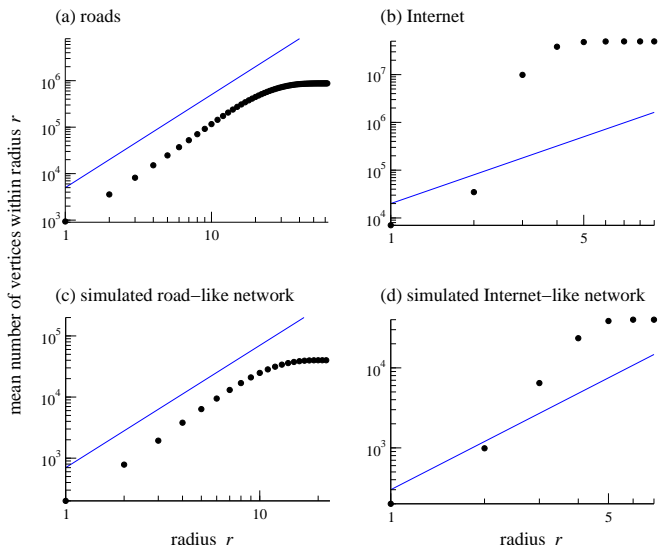}}
\end{center}
\caption{The size of neighborhoods vs.\ their radius on doubly-logarithmic 
plots (a)~for interstate highways, (b)~for the Internet, (c) and (d) for
simulations based on the optimization model described in the text.  The
straight lines have slope $2$ and indicate the expected growth for
two-dimensional networks.}
\label{dimplot}
\end{figure}

On an infinite regular $d$-dimensional lattice, such as a square or cubic
lattice, the dimension~$d$ can be calculated from $d=\lim_{r\to\infty}
d\log N_v(r)/d\log r$ where $N_v(r)$ is the number of vertices $r$ steps or
less from a given vertex~$v$~\cite{NW99b,CS04}.  On finite lattices one
cannot take the limit $r\to\infty$, but good results for~$d$ can be
achieved by plotting $\log N_v$ against $\log r$ for some central
vertex~$v$ and measuring the slope of the initial part of the resulting
line.  This idea can be used also to define an effective dimension for
networks.  (In order to reduce statistical errors, $N_v$~is averaged over
all vertices~$v$, but in other respects the calculation is identical.)  We
show the resulting plots for the interstate network and the Internet in
Fig.~\ref{dimplot}, panels (a) and (b).  As the figure shows, the slope of
the plot is close to~$2$ for the interstates, indicating that this network
is essentially two-dimensional.  For the Internet on the other hand, the
plot grows much faster with~$r$, indicating that the network has high
dimension, or perhaps no well-defined dimension at all (similar results are
seen for the airline network).

If a network is fundamentally two-dimensional, then we would expect it to
have a diameter that, like any two-dimensional system, varies as the square
root of the network size.  Essentially all other networks, by contrast,
have diameters varying much more slowly, usually logarithmically with
network size.  Thus, we propose a tentative explanation of the structure of
our geographic networks as follows.  All the networks appear to show a
preference for short edges over long ones, which is a natural effect of
geography.  However, the road network has much shorter edges, lower
degrees, and larger diameter than the other two.  These are all expected
consequences of a two-dimensional or planar form, and when we measure
dimension we do indeed find that the road network is fundamentally
two-dimensional, while the other networks are not.

This is a satisfying finding, certainly, but to some extent it just passes
the intellectual buck: our measurements can be explained in terms of
network dimensionality, but why do the networks have different dimension in
the first place?  As we now show, it is possible to construct a simple
model that explains the basic features of geographic networks, including
their dimension, in terms of competing preferences for either short
Euclidean distances between vertices or short graph distances.

First, let us assume that the cost of building and maintaining a network is
proportional to the total length of all its edges:
\begin{equation}
\text{cost} = \!\sum_{\text{edges\ }(i,j)}\!\!d_{ij},
\label{cost}
\end{equation}
where $d_{ij}$ is the Euclidean length of the edge between vertices $i$
and~$j$.  This result is only approximately true in most cases, but it is a
plausible starting point.

From a user's perspective, a network will usually be better if the paths
between points are shorter.  As we have seen, however, the way we measure
path length can vary.  In a road network most travelers look for routes
that are short in terms of miles, while for airline travelers the number of
legs is often considered more important.  To account for these differences,
we assign to each edge an effective length thus:
\begin{equation}
\mbox{effective length of edge $(i,j)$}
  = \lambda\sqrt{n}\>d_{ij} + (1-\lambda),
\end{equation}
where $0\le\lambda\le1$ and $n$ is the number of vertices.  The parameter
$\lambda$ determines the user's preference for measuring distance in terms
of miles or legs.  (The factor of $\sqrt{n}$ is not strictly necessary but
it is convenient; it compensates for the scaling of nearest-neighbor
distances $d_{ij}\sim n^{-1/2}$ with system size.)  Now we define the total
distance between two (not necessarily adjacent) vertices to be the sum of
the effective lengths of all the edges along a path between them, minimized
over all paths.

\begin{figure}
\centering
\resizebox{8cm}{!}{\includegraphics{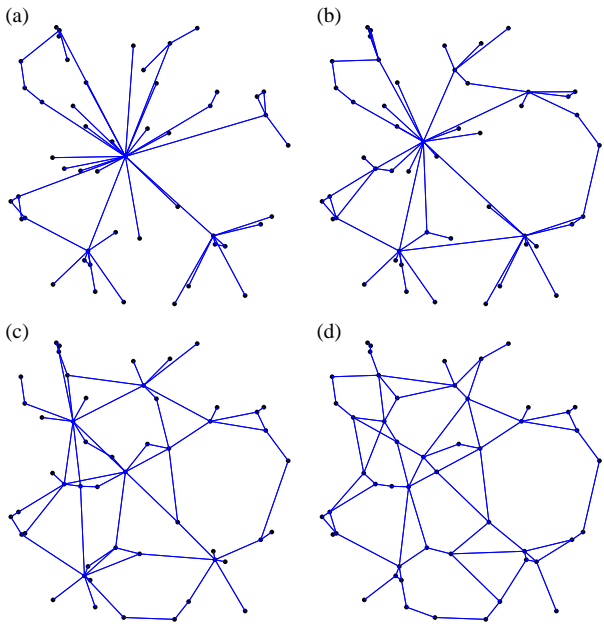}}
\caption{Optimized network structures for (a)~$\lambda=0$,
(b)~$\lambda=\frac13$, (c)~$\lambda=\frac23$, (d)~$\lambda=1$.  Networks
(a) and (d) resemble airline and road networks respectively, while (b) and
(c) show structure intermediate between the two extremes.}
\label{model}
\end{figure}

We now construct a model network as follows.  We suppose we are given the
positions of $n$ vertices that we are to connect, we are given a budget,
Eq.~\eref{cost}, for building the network, and we are given the preference
of the users, meaning we are given a value of~$\lambda$.  We then search
for network structures that connect all the vertices, can be built within
budget, and minimize the mean vertex--vertex distance between all vertex
pairs, for edge lengths defined as above.  This is a standard combinatorial
optimization problem, for which we can derive good (though usually not
perfect) solutions using simulated annealing.

Fig.~\ref{model} shows four networks generated in this fashion for $n=50$
vertices placed at random within a square.  For $\lambda=0$ and $\lambda=1$
we find networks strongly reminiscent of airlines and roads
respectively---tree-like structures with long edges and hubs in the first
case and structures with neither long edges nor hubs in the second.  For
intermediate values of $\lambda$ the model finds a compromise between hub
formation and local links.

To make this comparison more concrete, we have also generated networks with
the same mean degrees as our three empirically observed networks.  For
$n=200$ nodes, we find that the maximum degree of the model networks varies
between $7$ ($3.5\%$ of the network) and $143$ ($71.5\%$) as we vary
$\lambda$ from $0$ to~$1$.  At the same time, the diameter decreases from a
sizable $21$ to a small-world-like~$4$.  In Fig.~\ref{dimplot}c
and~\ref{dimplot}d we show the mean size of the neighborhood $N_v(r)$ of a
vertex as a function of distance~$r$, as we did for our empirically
observed networks.  As the figure shows, the results indicate a network
with a roughly two-dimensional form for large~$\lambda$
(Fig.~\ref{dimplot}c) and a strongly super-quadratic form for
small~$\lambda$ (Fig.~\ref{dimplot}d).  All of these results are in
excellent agreement with our empirical observations for the real airline
and road networks.

We propose therefore that the qualitative features of spatial networks can
be well represented by a simple one-parameter family of networks balancing
miles traveled with number of legs between vertex pairs.  Typical road
networks have the structure one would expect if their users care primarily
about the length of their journey in miles, while airline networks
correspond to users who care primarily about minimizing the number of legs.

The results presented here are, inevitably, only the beginnings of a
detailed study of spatial networks.  Many other features of these networks
deserve scrutiny, such as, for instance, the effects of population
distribution.  We hope that others will also investigate this interesting
class of systems and look forward with anticipation to their results.

The authors thank the staff of the University of Michigan's Numeric and
Spatial Data Services for their help with the geographic data.  This work
was funded in part by the National Science Foundation under grant number
DMS--0234188 and by the James S. McDonnell Foundation.

\end{document}